\documentclass[
reprint,
superscriptaddress,
showpacs,
nofootinbib,
amsmath,amssymb,
aps,
pra,
floatfix,longbibliography
]{revtex4-2}
\usepackage{graphicx}
\usepackage{xeCJK}
\usepackage{dcolumn}
\usepackage{booktabs}
\usepackage{amsmath,amssymb,amsfonts,amsthm}
\usepackage{mathtools}
\usepackage{txfonts}
\usepackage{bbm}
\usepackage{bm}
\usepackage{mathrsfs}
\usepackage{xcolor}
\usepackage{siunitx}

\newcommand{\ket}[1]{|#1\rangle}

\usepackage{makecell}
\usepackage[hypertexnames, breaklinks=true, bookmarksnumbered=true, bookmarksopen=true, colorlinks=true, linktocpage=true, citecolor=blue, urlcolor=magenta, linkcolor=magenta]{hyperref}
\begin{document}

\title{Parametrically Driven Superradiance of an Interacting Tavis-Cummings Model}

\author{Wen-Jie Geng(耿文杰)}
  \affiliation{Laboratory of Quantum Information, University of Science and Technology of China, Hefei 230026, China}
 \author{Yiwen Han(韩奕文)}
  \affiliation{Laboratory of Quantum Information, University of Science and Technology of China, Hefei 230026, China}
\author{Wei Yi(易为)}
\email{wyiz@ustc.edu.cn}
 \affiliation{Laboratory of Quantum Information, University of Science and Technology of China, Hefei 230026, China}
\affiliation{Anhui Province Key Laboratory of Quantum Network, University of Science and Technology of China, Hefei 230026, China}
\affiliation{CAS Center For Excellence in Quantum Information and Quantum Physics, Hefei 230026, China}
\affiliation{Hefei National Laboratory, University of Science and Technology of China, Hefei 230088, China}
\affiliation{Anhui Center for Fundamental Sciences in Theoretical Physics, University of Science and Technology of China, Hefei 230026, China}

\begin{abstract}
We consider the superradiant transition of a generalized Tavis-Cummings model, where a number of two-level qubits are coupled to a dissipative cavity. The cavity is coherently driven through a parametric medium, and all-to-all interactions between the qubits are introduced.
While the nonlinear gain from the parametric drive breaks the $U(1)$ symmetry of the standard Tavis-Cummings model, thus giving rise to superradiance with squeezed cavity fields, we show that the interactions impact the collective excitations and significantly modify the superradiant transition.
Insights to the superradiant phase transitions, as well as the interaction effects, are obtained through effective models involving only a handful of low-lying collective states, under which the steady-state phase diagram of the hybrid system is faithfully reproduced.
Our study is relevant to Rydberg-atom arrays coupled to a parametrically driven cavity, where the long-range interactions derive from the dipole-dipole interatomic interactions.
\end{abstract}

\maketitle

\section{Introduction}

Superradiance is a signature phenomenon in quantum optics and cavity quantum electrodynamics, wherein an ensemble of atoms are coupled to a quantized cavity field~\cite{PhysRevA.3.1735,PhysicsReports.93.301,Phys.Rev.Lett.92.073602,PhysRevA.83.033601,Front.Phys.13.136701,arXiv:2509.04391}. Owing to the collective nature of the atom-cavity coupling, the hybrid system undergoes a phase transition to become superradiant when the light-matter coupling exceeds a threshold~\cite{Phys.Rev.A.8.1440,AnnalsofPhysics.76.360,PhysicsLettersA46.47,Phys.Rev.A.10.1437,Phil.Trans.R.Soc.A.369.1137,PLoS.ONE.15.e0235197}.
In the superradiant phase, the cavity field is macroscopically populated and collective excitations emerge in the atoms. In the normal phase below the threshold, the cavity field vanishes and atoms remain in their ground states.
Superradiance and the superradiant phase transition offer a fundamentally important paradigm for collective quantum phenomena as well as quantum phase transitions in both closed and open systems.
They have therefore stimulated extensive interest in the past decades~\cite{Phys.Rev.A.98.023804,PhysRevA.87.053623,Phys.Rev.A.87.023831,PhysRevLett.90.044101,Adv.QuantumTechnol.2.1800043,Sci.Bull.63.542,ChinesePhys.Lett.41.064201,ChinesePhys.B33.054204}.

An exemplary description for the superradiant transition is provided by the Dicke model~\cite{Phys.Rev.A.7.831,Phys.Rev.A.102.023703,Phys.Rev.A.95.053854,NewJ.Phys.20.013006,Phys.Rev.Research.6.033181,Nature.464.1301,J.Phys.B:At.Mol.Opt.Phys.39.3315}, where a number of qubits (or two-level atoms) couple to a single cavity mode. The counter-rotating terms therein help to break the continuous $U(1)$ symmetry, facilitating the onset of superradiance where a $Z_2$ discrete symmetry is spontaneously broken. This is in contrast to the Tavis-Cummings model~\cite{Phys.Scr.79.065405,Nature.543.87,IEEETrans.Automat.Contr.68.2048,PhysRevA.85.043815,NewJ.Phys.15.123032,PhysRevA.94.033808,PhysRevA.85.053621,J.Phys.A:Math.Gen.29.6305}, which preserves the $U(1)$ symmetry and prohibits the superradiant transition.
In a recent study~\cite{PhysRevletters.124.073602}, it is shown that a parametrically driven Tavis-Cummings model can undergo the superradiant phase transition, due to the introduction of nonlinear parametric terms that break the $U(1)$ symmetry of the standard Tavis-Cummings model.
It follows that the phase transition features a $Z_2$ symmetry breaking, accompanied by the superradiant generation of squeezed light.

The occurrence of superradiance sensitively depends on the collective excitations in the atomic sector. For instance, the Dicke states, a class of collective states with superposed local excitations, play a dominant role across the superradiant transition~\cite{Hyperfine.Interact.37.71}. Modifying the collective excitations, either through quantum statistics or engineered interactions, can enhance the superradiance~\cite{PhysRevletters.112.143004,PhysRevLett.133.243401,PhysRevLett.112.143002,Science.373.1359,PhysRevLett.112.143003}, so that the system becomes superradiant under an infinitesimally small coupling strength. Intuitively, related phenomena of similar mechanism should also arise in the parametrically driven Tavis-Cummings model.

In this work, we investigate in detail the impact of interactions on the superradiance of a parametrically driven dissipative Tavis-Cummings model. The parametric drive introduces quadratic terms in the cavity field, breaking the $U(1)$ symmetry of the standard Tavis-Cummings model and leading to the superradiance of squeezed light.
The system under study can be implemented using Rydberg-atom arrays in a cavity, where the interatomic interactions originate from dipole-dipole interactions between Rydberg atoms~\cite{PhysRevA.106.L021101,PhysRevA.105.L041302,J.Phys.B:At.Mol.Opt.Phys.49.064014,PhysRevB.106.134506,Nat.Phys.19.714}. While the dipolar interactions feature a $1/R^3$ tail (with $R$ the interatomic distance), for convenience, we focus on the case with homogeneous all-to-all interactions. Using a mean-field approach~\cite{Sci.Rep.10.13408,Phys.Rev.Lett.126.230601,Proc.R.Soc.A.474.20170856,Phys.Rev.A107.033711,Phys.Scr.90.068015,Phys.Rev.B.106.245124}, we show that the superradiant transition is dependent not only on the atom-cavity coupling, but also on the parametric drive and the interaction strength. In the superradiant state, the cavity field and the atomic excitations can occupy either of two degenerate states, reflecting the $Z_2$-symmetry breaking of the transition~\cite{Phys.Rev.A.98.042118}.
Consistent with the absence of superradiance in a standard Tavis-Cummings model, superradiance can only emerge when the nonlinear gain from the parametric drive exceeds the cavity decay.
Importantly, we find that repulsive interactions tend to suppress superradiance, making it occur at a larger parametric drive.
On the other hand, while attractive interactions generally enhance superradiance, near a discrete set of attractive interaction strengths, the superradiant phase boundary becomes nonmonotonic in the parametric drive.
These fine structures on the attractive-interaction side vanish for sufficiently large atom-cavity coupling.
Using an effective model with four lowest-lying Dicke states, we are able to faithfully reproduce the steady-state phase diagram, and show that the interaction effects can be qualitatively captured by considering the energy gap between the two lowest-lying Dicke states. This suggests that these two states play a dominant role in the superradiant transition.
We then analytically derive the phase boundaries in the thermodynamic limit with the atom number $N\to \infty$, and discuss the evolution of the phase boundary with increasing $N$.
Our work highlights the interplay of parametric drive, cavity decay, and many-body interactions on the superradiant phase transition, with results readily accessible by coupling Rydberg-atom arrays to a cavity.

The work is organized as follows. In Sec.~II, we introduce the model and discuss its potential implementation using Rydberg atoms. We then study the dynamics and steady-state phase diagrams in the absence of interactions in Sec.~III. The impact of interactions on the phase diagram is then investigated in Sec.~IV, where an effective model consisting of a handful of Dicke states is shown to be capable of reproducing the phase diagram. We also study in detail the finite-size effects in this section, and demonstrate that the impact of interactions is accessible under typical experimental conditions. Finally, we summarize in Sec.~V.

\begin{figure}
    \centering
   \includegraphics[width=0.9\linewidth]{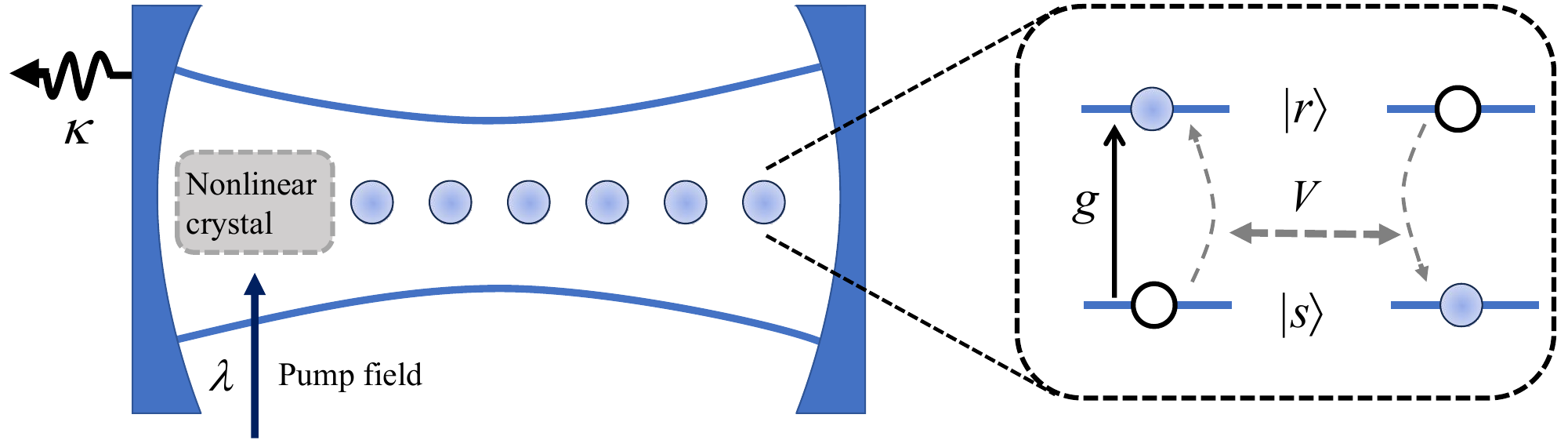}
    \caption{Schematic of the parametrically driven, dissipative Tavis-Cummings model.
     The cavity contains a nonlinear crystal pumped by a laser field.
     Each qubit, in the basis of $\{|s\rangle, |r\rangle\}$, is driven by the cavity mode with a coupling rate $g$. All-to-all interactions with strength $V$ exist between the qubits. The cavity decay rate is given by $\kappa$.
    }
    \label{fig1}
    \end{figure}

\section{Model}

As illustrated in Fig.~\ref{fig1}, we consider a number of $N$ qubits coupled to a single cavity mode, which is in turn generated through an optical parametric amplification process~\cite{Phys.Rev.Lett.120.093602,Phys.Rev.A80.033807,Phys.Rev.Lett.95.233601}. Taking the cavity decay into account, the dynamics of the system are described by the Lindblad master equation
\begin{equation}
    \frac{d}{dt}\hat{\rho} = -i \left[ \hat{H}_{\text{tot}}, \hat{\rho} \right] + \kappa \mathcal{D}[\hat{a}],
    \label{eq:master}
\end{equation}
where $\hat{\rho}$ is the density matrix of the qubit-cavity hybrid system, $\kappa$ is the cavity decay rate, $\mathcal{D}[\hat{a}]=2\hat{a}\rho\hat{a}^{\dagger}-\{\hat{a}^{\dagger}\hat{a},\rho\}$, and $\hat{a}$ ($\hat{a}^\dagger$) is the annihilation (creation) operator of the cavity field.

The coherent Hamiltonian $H_{\text{tot}}$ reads
\begin{align}
\hat{H}_{\text{tot}} &= \Delta_{c}\hat{a}^{\dagger}\hat{a}+\Delta_{a}\sum_{i=1}^N\frac{\sigma_{i}^{z}}{2}+\frac{g}{\sqrt{N}}\sum_{i=1}^N(\hat{a}^{\dagger}\hat{\sigma}_{i}^{-}+\hat{a}\hat{\sigma}_{i}^{+})\nonumber\\
&+\frac{\lambda}{2}(\hat{a}^{\dagger2}+\hat{a}^{2})+\sum_{i<j}\frac{V}{N}(\hat{\sigma}_{i}^{+}\hat{\sigma}_{j}^{-}+\text{H.c.}),
\label{eq2}
\end{align}
where the cavity and qubit detunings are respectively $\Delta_c=\omega_c-\omega_{L}$ and $\Delta_a=\omega_a-\omega_{L}$, with $\omega_c$, $\omega_a$, and $\omega_L$ the frequencies
of the cavity mode, the qubit, and the pumping field, respectively. The ladder operators $\hat{\sigma}^{\pm}_i=(\hat{\sigma}^x_i\pm i \hat{\sigma}^y_i)/2$ are defined through the Pauli operators $\hat{\sigma}^{x,y,z}_{i}$ of the $i$th qubit.
While the parameter $g$ characterizes the qubit-cavity coupling, $\lambda$ is the nonlinear gain coefficient resulting from the parametric amplification, which is proportional to the amplitude of the pumping field. The corresponding contribution to the Hamiltonian, quadratic in the cavity field, leads to the squeezing of the cavity mode. The last term of Hamiltonian (\ref{eq2}) describes the all-to-all interactions, with a uniform interaction strength $V$, between the qubits. Throughout this work, we take $\Delta_{a}=\Delta_{c}=\Delta=1$ as the unit of energy.

Such a hybrid configuration can in principle be implemented using Rydberg-atom arrays in a cavity, pumped through a parametric medium. The two internal states of each qubit correspond to two different Rydberg states, labeled respectively as $|s\rangle$ and $|r\rangle$ in Fig.~\ref{fig1}, with dipole-dipole interactions in between. While the inter-atomic interactions therein feature a $1/R^3$ tail, we focus on the simplified case with all-to-all interactions for convenience. We expect that the key properties of the system remain qualitatively similar when the interaction range is properly taken into account~\cite{PhysRevLett.133.243401}.

To proceed, we introduce the collective operators $\hat{J}_{z}=\sum_{i=1}^N\frac{\hat{\sigma}_i^{z}}{2}$, $\hat{J}_{+}=\sum_{i=1}^N\hat{\sigma}_i^{+}$, $\hat{J}_{-}=\sum_{i=1}^N\hat{\sigma}_i^{-}$, so that the coherent Hamiltonian can be rewritten in a more compact form
\begin{align}
\hat{H}_{\text{tot}} &= \Delta_{c}\hat{a}^{\dagger}\hat{a}+\Delta_{a}\hat{J}_{z}+\frac{g}{\sqrt{N}}(\hat{a}^{\dagger}\hat{J}_-+\hat{a}\hat{J_{+}})+\frac{\lambda}{2}(\hat{a}^{\dagger2}+\hat{a}^{2})\nonumber\\
&+\frac{V}{2N}(\hat{J}_{+}\hat{J}_{-}+\hat{J}_{-}\hat{J}_{+}-N).
\label{eq3}
\end{align}

From the Heisenberg equations of motion, we have
\begin{align}
i\dot{\hat{a}} &=(\Delta_c-i\kappa)\hat{a}+\frac{g}{\sqrt{N}}\hat{J}_{-}+\lambda\hat{a}^{\dagger},\label{eq4}\\
i\dot{\hat{J}}_{-} &=(\Delta_a+\frac{V}{N})\hat{J}_{-}-\frac{2 g }{\sqrt{N}}\hat{a}\hat{J}_{z}-\frac{2V}{N}\hat{J}_{-}\hat{J}_{z},\label{eq5}\\
i\dot{\hat{J}}_{z} &=\frac{g}{\sqrt{N}}(-\hat{a}^{\dagger}\hat{J}_{-}+\hat{a}\hat{J}_{+}).
\label{eq6}
\end{align}
In the following, we study the steady-state phase diagram of the system based on the equations of motion above.

\section{Superradiant phase transition for $V=0$}

We are interested in the steady-state solution of the system, which, in the absence of interactions, can be analytically solved under the mean-field approximation.
Specifically, replacing $\hat{a}$ with $\langle\hat{a}\rangle:=\alpha$, and imposing the stationary condition of Eq.~(\ref{eq4}), we have
\begin{align}
\alpha
=\frac{g\left[(\Delta_{c}+i\kappa)\langle\hat{J}_{-}\rangle-\lambda\langle\hat{J}_{+}\rangle\right]}{\sqrt{N}(\lambda^{2}-\Delta_{c}^{2}-\kappa^{2})}.
\label{eq7}
\end{align}
Note that, while the mean-field approximation becomes exact in the thermodynamic limit $N\rightarrow \infty$ \cite{Phys.Rev.Lett.126.230601}, it already provides a qualitatively correct picture for the interaction effects for a handful of atoms (for instance, $N\sim 10$).

Following Ref.~\cite{PhysRevletters.124.073602}, by taking $V=0$, the superradiant phase is simultaneously confined by two boundaries: $\lambda=\kappa$ and $\lambda=\sqrt{\kappa^2+(g^2-\Delta_a\Delta_c)^2/\Delta_a^2}$, which intersect at $\lambda=\kappa$ and $g=\sqrt{\Delta_a\Delta_c}$. Notably, for $g>\sqrt{\Delta_a\Delta_c}$, the phase transition uniformly occurs at $\lambda=\kappa$, and the cavity becomes superradiant as soon as the parametric pump exceeds the cavity decay. More specifically, for $g>\sqrt{\Delta_a\Delta_c}$, the boundary $\lambda=\sqrt{\kappa^2+(g^2-\Delta_a\Delta_c)^2/\Delta_a^2}$ separates the superradiant phase into two regions in terms of the number of the solutions. To the left of the boundary, the system hosts two degenerate superradiant steady states, whereas to the right, an additional normal-state solution coexists\cite{PhysRevletters.124.073602}. Owing to the divergence of Eq.~(\ref{eq7}) at $\lambda=\sqrt{\kappa^2+\Delta_{c}^2}$, we work in the regime $\lambda<\sqrt{\kappa^2+\Delta_{c}^2}$.
\begin{figure}
    \centering   \includegraphics[width=0.9\linewidth]{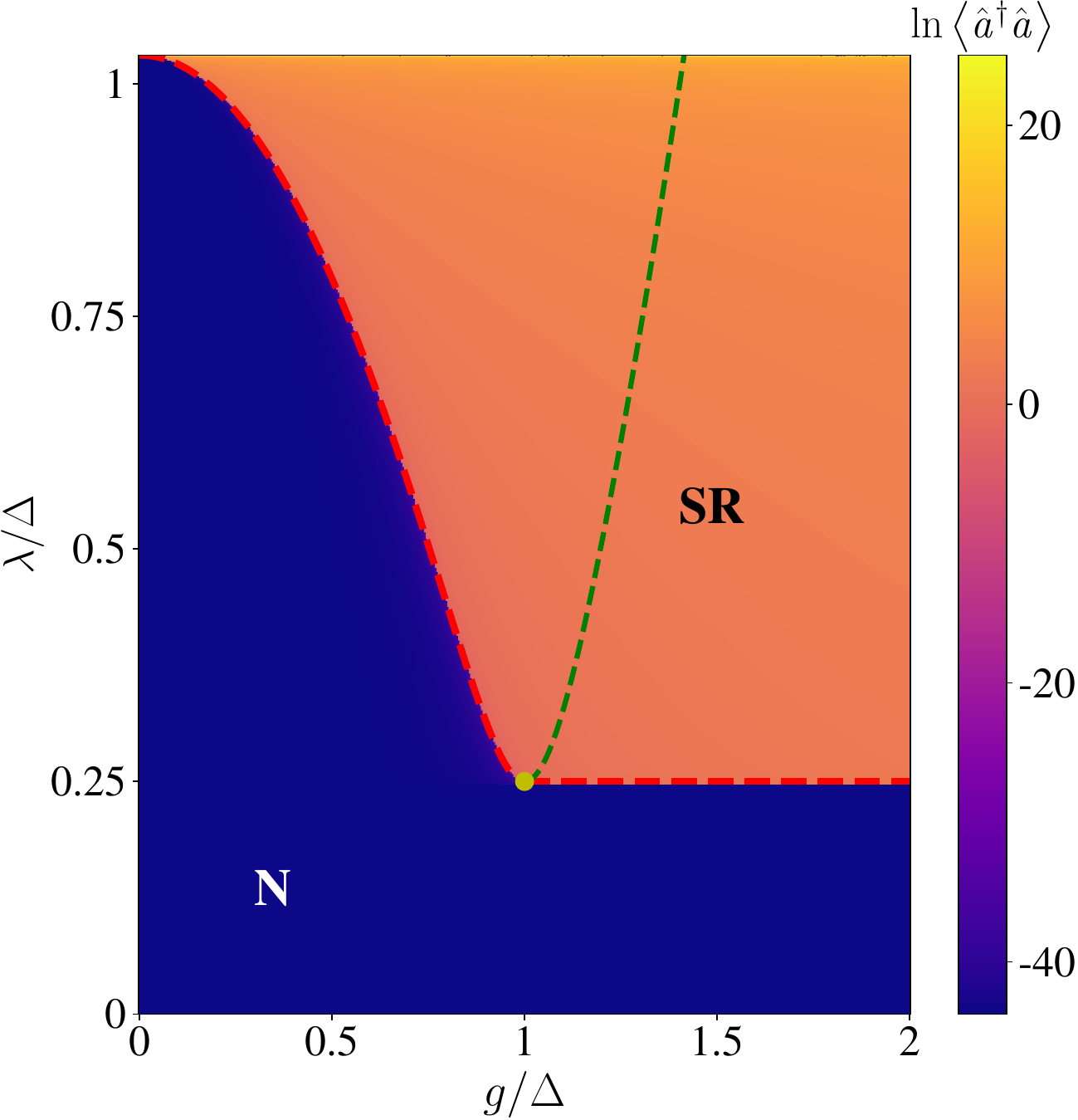}
    \caption{Steady-state phase diagram for $V=0$, characterized by the average cavity photon number. Here ``SR'' and ``N'' stand for the superradiant and normal phases, respectively. The red dashed curves comprise the analytically determined phase boundaries, below which the steady state of the system remains in the normal state. The yellow dot at $\lambda=\kappa$ and $g=\sqrt{\Delta_{a}\Delta_{c}}$ denotes the intersection of the two analytic boundaries (see main text). The green dashed line separates the superradiant phase into two regions: that with a pair of superradiant steady-state solutions to its left, and the one with coexisting superradiant and normal solutions to the right.
Other parameters are $\Delta_{a}=\Delta_{c}=1$, $\kappa=0.25$, and $N=8$.}
    \label{fig2}
    \end{figure}

\begin{figure}
    \centering
   \includegraphics[width=0.9\linewidth]{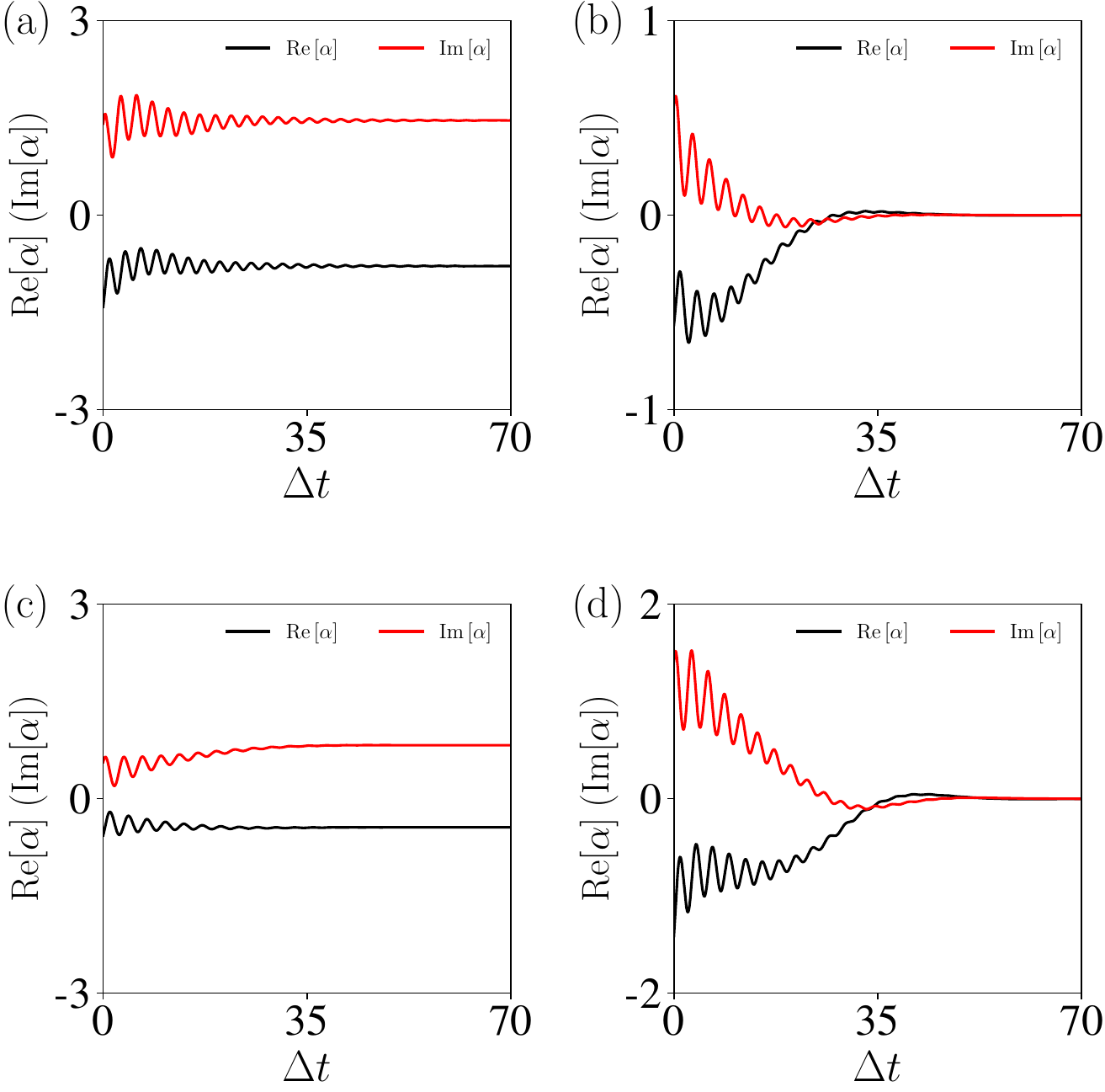}
    \caption{(a)(b) Time evolutions of the cavity mean field $\alpha$ in the superradiant phase with coexisting solutions.  We take the same parameters $g=1.2$, $\lambda=0.3$ in (a)(b), but different initial states. (c)(d) Dynamics of the cavity mean field $\alpha$ for regions with only the superradiant and normal solutions, respectively. We take
    $g=1$, $\lambda=0.3$ in (c), and $g=1.2$, $\lambda=0.2$ in (d). The black (red) line stands for the real (imaginary) components of $\alpha$.
Other parameters are $\Delta_{a}=\Delta_{c}=1$, $\kappa=0.25$, and $N=8$.}
    \label{fig3}
    \end{figure}

In Fig.~\ref{fig2}, we numerically determine the steady-state phase diagram by diagonalizing
Hamiltonian (\ref{eq3}) for its ground state while self-consistently imposing the stationary condition Eq.~(\ref{eq7}). The phase boundary is clearly visible from the color contour of the numerically evaluated steady-state photon number ($\text{ln}\langle \hat{a}^\dag\hat{a}\rangle$).
The numerical results for $N=8$ atoms are consistent with the analytical phase boundary (red dashed curve).
Note that the green dashed curve is the boundary of $\lambda=\sqrt{\kappa^2+(g^2-\Delta_a\Delta_c)^2/\Delta_a^2}$ in the region $g>\sqrt{\Delta_a\Delta_c}$, as discussed above.

To corroborate these results, we numerically evolve Eqs.~(\ref{eq4})(\ref{eq5})(\ref{eq6}), and plot the dynamics of $ \text{Re}[\alpha]$ and $\text{Im}[\alpha]$ in regions with coexisting
superradiant and normal solutions [see Fig.~\ref{fig3}(a)(b)], with only the superradiant solution  [see Fig.~\ref{fig3}(c)], and with only the normal solution [see Fig.~\ref{fig3}(d)], respectively. Notably, the superradiant solution in Fig.~\ref{fig3}(a) and the normal solution in Fig.~\ref{fig3}(b) coexist under the same parameters, with $g=1.2$ and $\lambda=0.3$.
While the steady-state solutions are consistent with the phase diagram in Fig.~\ref{fig3}(a)(c), the complex nature of $\alpha$ is a
combined result of cavity decay and the quadratic term from the parametric drive.


\begin{figure}
    \centering
   \includegraphics[width=0.9\linewidth]{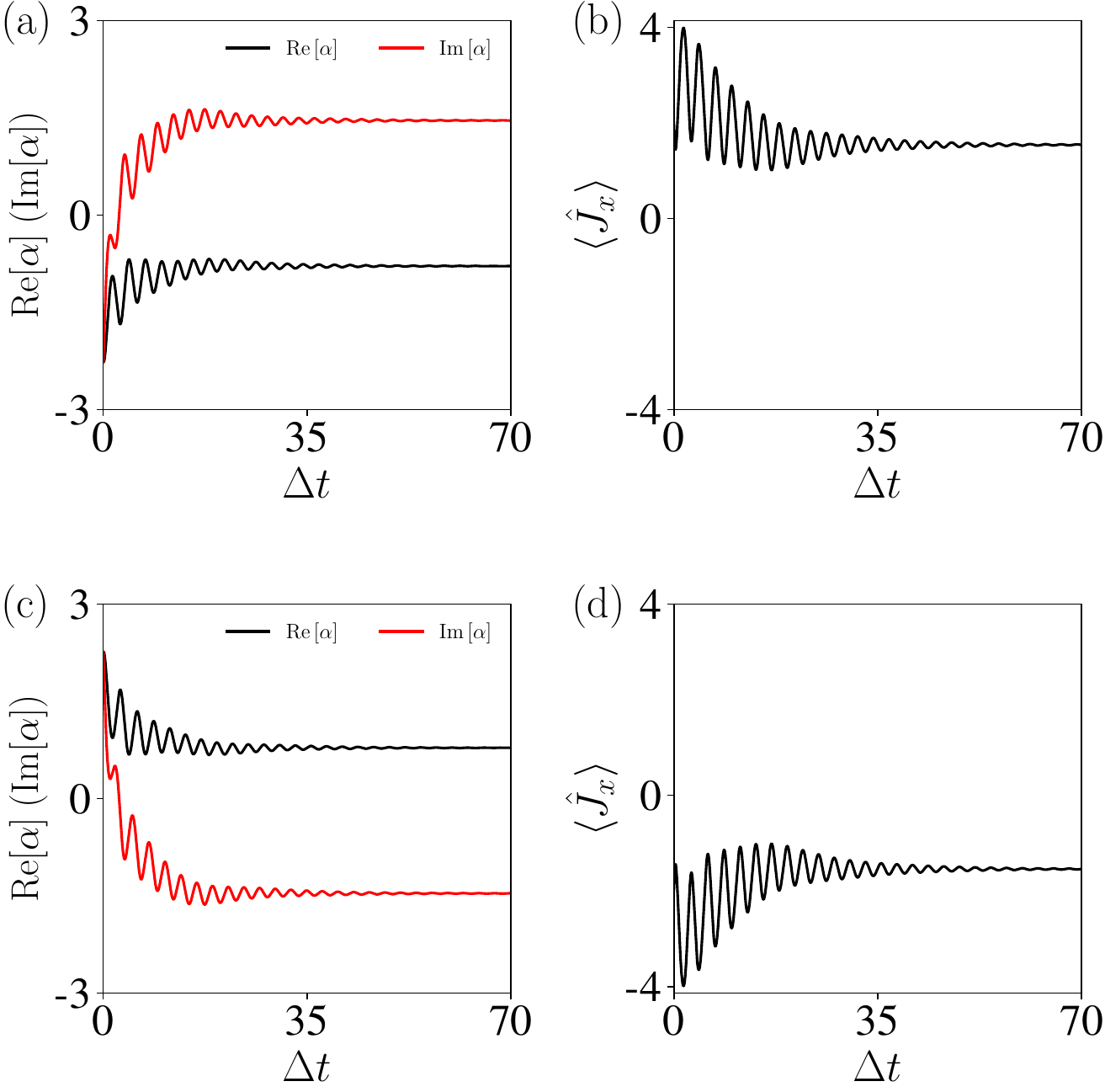}
    \caption{(a)(c) Time evolutions of $\alpha$ for $\lambda=0.3$ and $g=1.2$, where the black (red) line represents the real (imaginary) components of $\alpha$.
    The steady states show different values: $\alpha=-0.78+1.46i$ in (a) and $\alpha=0.78-1.46i$ in (c). Different initial states are adopted for (a) and (c).
    (b)(d) Time evolution of $\langle\hat{J}_{x}\rangle$ with $\lambda=0.3$ and $g=1.2$.
    The steady states exhibit different expectation values:
    $\langle\hat{J}_{x}\rangle =1.54$ for (b) and $\langle\hat{J}_{x}\rangle =-1.54$ for (d).
Other parameters are $\Delta_{a}=1$, $\Delta_{c}=1$, $\kappa=0.25$, and $N=8$.}
    \label{fig4}
    \end{figure}

Hamiltonian (\ref{eq3}) possesses the $Z_{2}$ symmetry, characterized by its invariance under the transformation: $\hat{J_{x}}\xrightarrow{}-\hat{J_{x}}$, $\hat{J_{y}}\xrightarrow{}-\hat{J_{y}}$, and $\hat{a}\xrightarrow{}-\hat{a}$.
The $Z_2$ symmetry is spontaneously broken upon the superradiant phase transition, such that the steady state of the system can be either one of a pair of degenerate superradiant states.
This is demonstrated in Fig.~\ref{fig4} through the equations of motion.
In Fig.~\ref{fig4}(a)(c), the time evolution of $\alpha$ is shown for $\lambda=0.3$ and $g=1.2$, where the black (red) line represents the real (imaginary) components of $\alpha$.
Comparing the long-time steady-state values of $\alpha$ in Fig.~\ref{fig4}(a) and (c), we identify a relative phase of $e^{-i\pi}$ between the two possible steady-state values.
Correspondingly, the long-time expectation value of $\langle \hat{J}_x\rangle$ in the atom sector also acquires opposite signs, as illustrated in Fig.~\ref{fig4}(b)(d).
Note that in the absence of the quadratic term (with $\lambda=0$), Hamiltonian (\ref{eq3}) is reduced to the $U(1)$ symmetric Tavis-Cummings model, where the superradiant phase transition is fully suppressed.

\section{The Superradiant phase transition under interactions}

\begin{figure}[tbp]

   \includegraphics[width=1\linewidth]{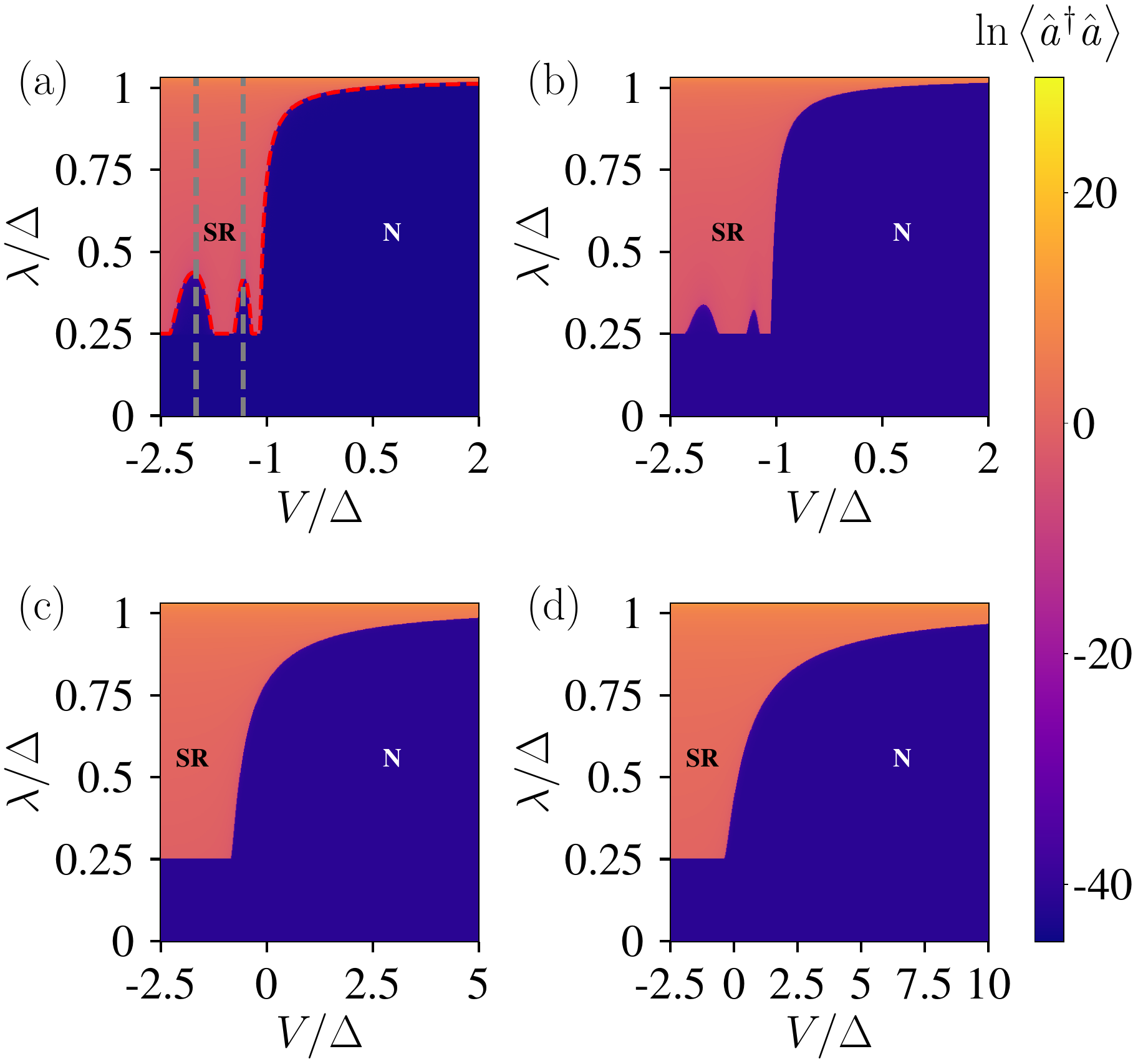}
    \caption{Steady-state phase diagrams with all-to-all interaction $V$, for (a) $g=0.2$, (b) $g=0.22$, (c) $g=0.5$, and (d) $g=0.8$, characterized by the average cavity photon number (color contour). The photon numbers are calculated using Hamiltonian (\ref{eq3}).
    The vertical gray dashed lines indicate the maximal suppression of superradiance. The red curve represents the phase boundary calculated using the effective Hamiltonian (\ref{eq9}).
   Other parameters are $\Delta_{a}=\Delta_{c}=1$, $\kappa=0.25$, and $N=8$.}
    \label{fig5}
    \end{figure}

We now consider the case with nonvanishing all-to-all interactions ($V\neq 0$).
In Fig.~\ref{fig5}, we plot the steady-state phase diagrams through the numerically calculated average photon number, for a finite number of atoms $N=8$ and with increasing atom-cavity coupling rate $g$.
Regardless of the value of $g$, repulsive interactions $V>0$ suppress the superradiant phase transition, pushing the phase boundary toward larger $\lambda$.
Here the phase boundary $\lambda_{c}$ increases monotonically with $V$ and asymptotically approaches a finite value for $V\to +\infty$.
On the other hand, while attractive interactions $V<0$ generally enhances superradiance, interesting fine structures in the phase boundary emerge
on the attractive-interaction side, where several additional normal regions protrude finger-like above the $\lambda=\kappa$ boundary, peaking at discrete interaction strengths.
Around these peaks, superradiance can be significantly suppressed.
Overall, the stability region of the normal phase decreases with increasing $g$, as the finger-like structures eventually disappear for sufficiently large $g$.

Qualitatively, the phase boundaries are modified since interactions impact the collective excitations in the atom sector. In fact, as we show below, the phase boundaries can be reproduced using a simplified effective model, consisting a handful of the collective states.
Motivated by the relevant collective states in the Dicke transitions, we consider an $N$-qubit Dicke state with $n$ excitations~\cite{Phys.Rev.Applied.044020}
\begin{align}
\ket{\psi_{n}} &=\frac{1}{\sqrt{C(N,n)}}\sum_{i} P_{i}\left[\ket{r}^{\otimes n}\otimes\ket{s}^{\otimes(N-n)}\right],
\label{eq8}
\end{align}
where $\sum_{i} P_{i}$ indicates the sum over all possible permutations of the qubit locations, and $C(N,n)$ is the combinatorial number. Projecting Hamiltonian (\ref{eq3}) into the subspace of Dicke states, we have
\begin{align}
\hat{H}_{\text{eff}} = &\sum_{n=0}^{d} \omega_n |\psi_n\rangle\langle\psi_n|+ \frac{g}{\sqrt{N}}\left(\hat{a}\sum_{n=0}^{d-1}\eta_n |\psi_{n+1}\rangle\langle\psi_n| + \text{H.c.}\right)\nonumber \\
+&\frac{\lambda}{2}(\hat{a}^2+\hat{a}^{\dagger2})+ \Delta_c a^\dagger a,
\label{eq9}
\end{align}
where $\omega_n=(n-N/2)\Delta_{a}+n(N-n)V/N$, $\eta_n=\sqrt{(N-n)(n+1)}$,
and $d+1$ is the number of lowest-lying Dicke states that we consider.

Crucially, we find that the phase boundaries can be reproduced for $d=3$, that is, by taking into account the lowest four Dicke states. This is illustrated in Fig.~\ref{fig4}(a), where the color contours in the background are numerically calculated using the Hamiltonian (\ref{eq3}), and the red dashed curve is the phase boundary calculated using the effective Hamiltonian (\ref{eq9}).
Notice that the normal-state finger tips are located at $V=-2$ and $V=-4/3$ (vertical gray dashed lines) for the parameter range that we consider.

\begin{figure}[tbp]
    \centering
   \includegraphics[width=0.9\linewidth]{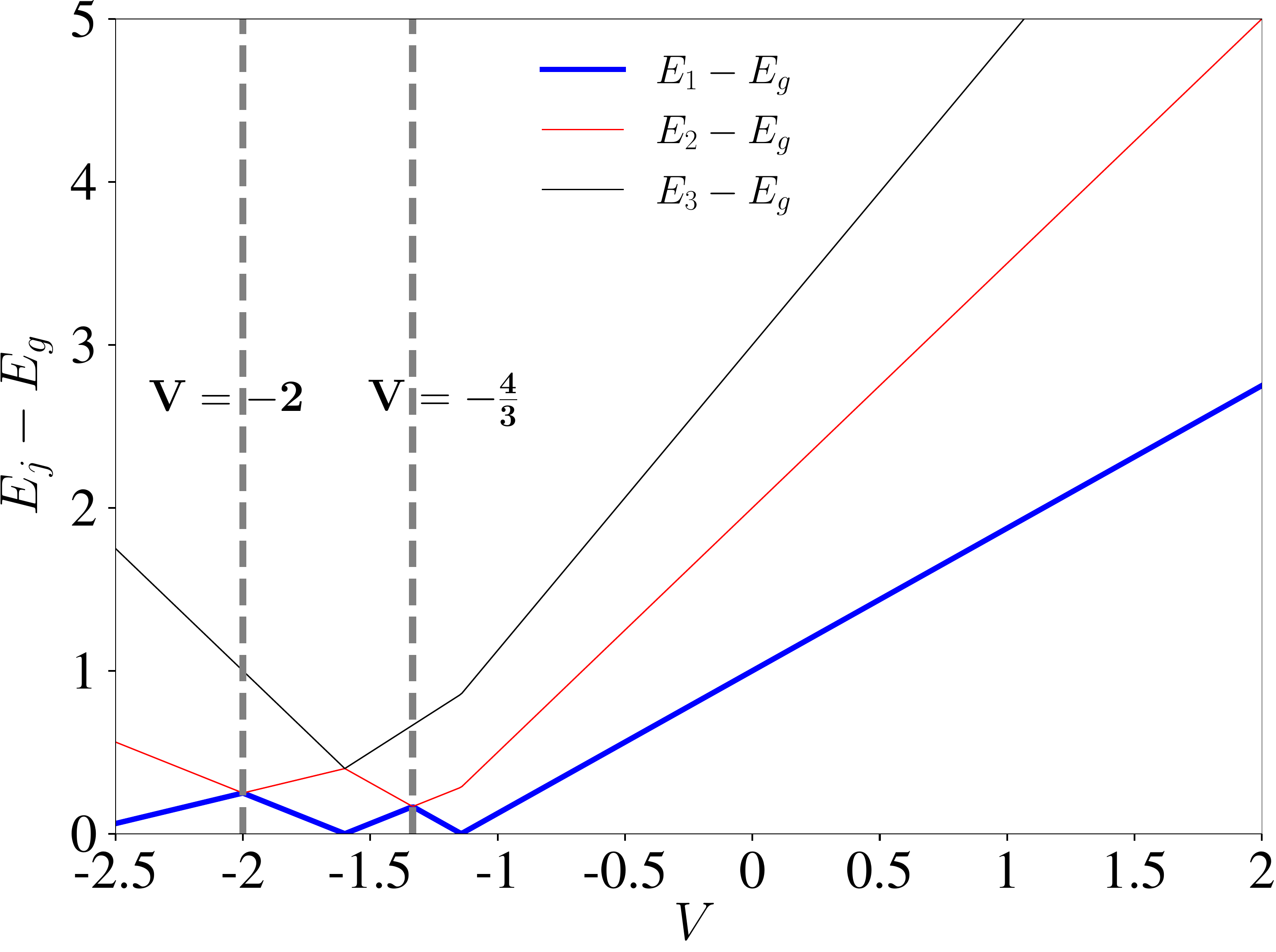}
    \caption{Energy gaps between the ground Dicke state and other low-lying Dicke states. Here $E_g$ is the energy of the ground Dicke state, and $E_{1,2,3}$ represent the energies of the first-, second-, and third-excited Dicke states, respectively. Other parameters are the same as those in Fig.~\ref{fig5}.}
    \label{fig6}
    \end{figure}

In Fig.~\ref{fig6}, we plot the energy gap between the lowest Dicke state (the ground state) and the other three states used in the effective model, respectively. On the attractive-interaction side, the positions where the superradiance is maximally suppressed (finger tips) coincide with the local maxima of the energy gap between the two lowest-lying Dicke states.
Recalling that, in a quantum Rabi model~\cite{Phys.Rev.A97.013825,PhysRevLett.133.243401}, the threshold of the superradiant transition increases with larger energy difference between the two energy levels, the phenomenon observed here can be understood in a similar vein. Apparently, near the superradiant phase transition here, the two lowest-lying collective states play a dominant role, though the contribution of the second- and third-excited states cannot be fully ignored.
Consistently, for $V>0$, the energy gap between the ground state and the first excited Dicke state increases monotonically with $V$, explaining the monotonic variation of phase boundary for repulsive interactions in Fig.~\ref{fig5}.


In Fig.~\ref{fig6}, we also notice two locations where the energy gap between ground state and first-excited state vanishes. One may expect that an interaction-enhanced superradiance here,
similar to the prediction in Ref.~\cite{PhysRevLett.133.243401} where an infinitesimally small $\lambda$ would lead to superradiance.
However, this would contradict with the absence of superradiance in a standard
Tavis-Cummings model. Indeed, in the parametrically driven Tavis-Cummings model, normal state persists up to $\lambda=\kappa$~\cite{PhysRevletters.124.073602}.


\begin{figure}[tbp]
    \centering
   \includegraphics[width=1\linewidth]{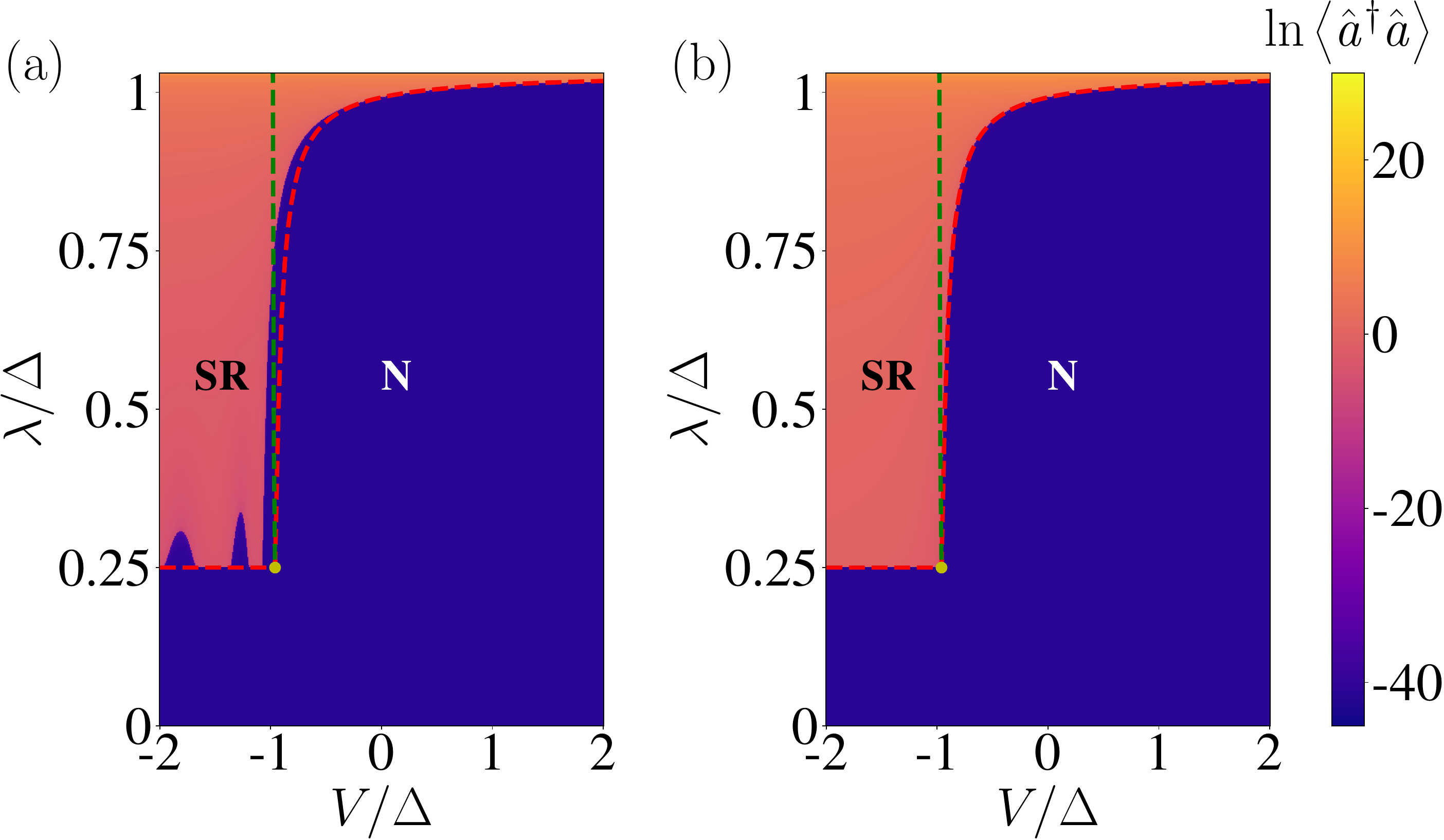}
    \caption{Steady-state phase diagrams for (a) 9 atoms and (b) 75 atoms, characterized by the average photon numbers (color contour). The red curves in (a)(b) indicate the phase boundaries in the thermodynamic limit with $N\to\infty$. The yellow dot is located at $V=-\Delta_{a}+\frac{g^2}{\Delta_c}$ and $\lambda=\kappa$ (see main text for discussions). The green curve separates the superradiant phase into two regions with different number of steady-state solutions (see also Fig.~\ref{fig2}). A pair of superradiant steady-state solutions exist to the right of the green curve, and superradiant and normal solutions coexist to the left.
    Other parameters are $\Delta_{a}=\Delta_{c}=1$, $g=0.2$, and $\kappa=0.25$.}
    \label{fig7}
    \end{figure}

Furthermore, we examine the effect of the total atom number $N$. This is motivated by the fact that the local
maxima of the energy gaps between Dicke states become smaller with increasing $N$. Hence, one expect that, above a certain $N$, the tips of the normal-state fingers would sink below $\lambda=\kappa$, as superradiance is effectively enhanced with larger $N$.
This is indeed observed in Fig.~\ref{fig7}, where we fix $g$ and demonstrate the steady-state phase diagrams for $N=9$ and $N=75$, respectively.

\begin{figure}
    \centering \includegraphics[width=1\linewidth]{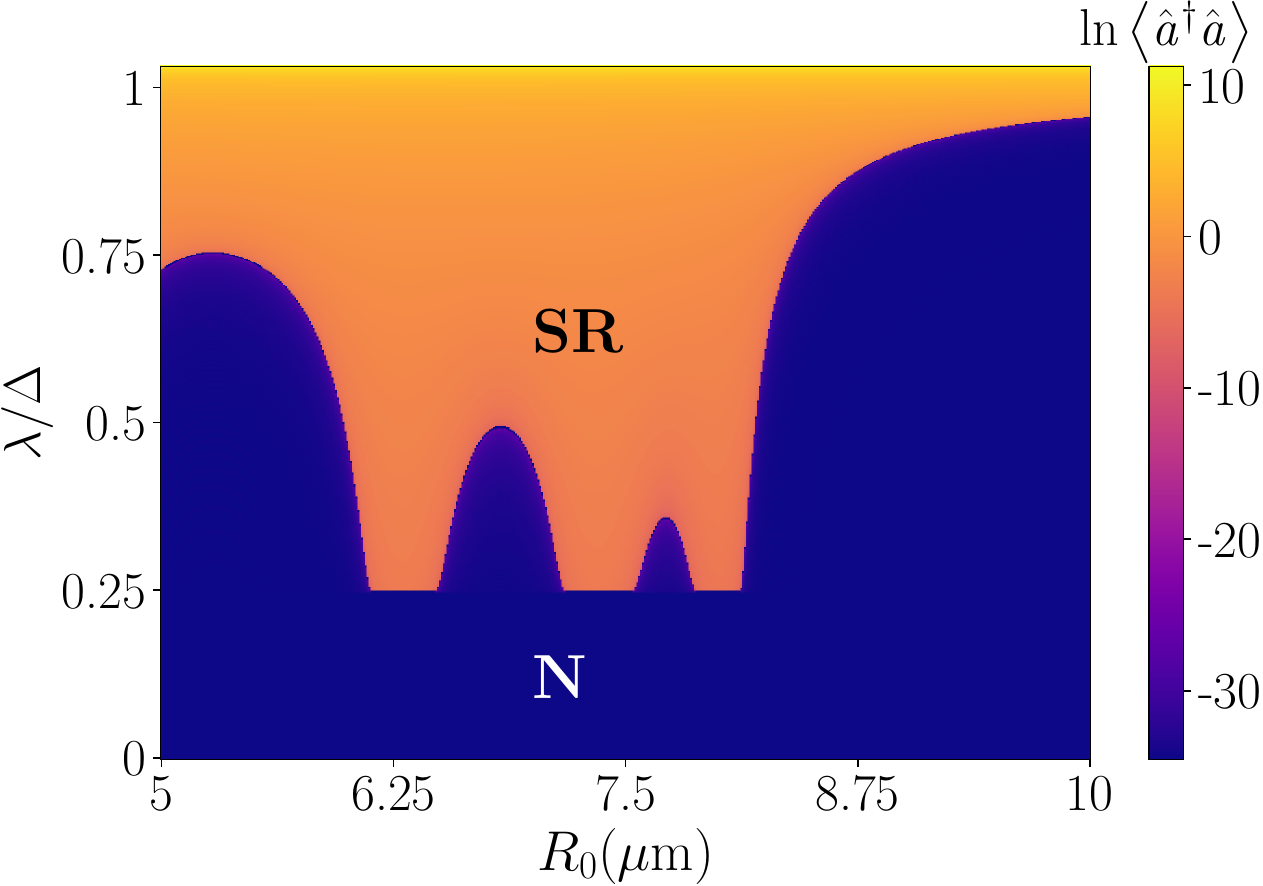}
    \caption{Steady-state phase diagram with realistic interaction $V_{\mathrm{r}}$.  We take the Rydberg states $\ket{60S_{1/2},m_{j}=\frac{1}{2}}$ and $\ket{60P_{1/2},m_{j}=-\frac{1}{2}}$ of $^{87}\rm{Rb}$ as $\ket{s}$ and $\ket{r}$, as well as the typical experimental parameters: $\Delta_c=\Delta_a=\Delta=2\pi\times6.3\mathrm{MHz}$, $\frac{C_{3}}{\hbar\Delta}=-240.48\mu m^3$, $\kappa/\Delta=0.25$, $g/\Delta=0.2$, and $N=8$~\cite{Nature.616.691,Nature.464.1301}.}
    \label{fig8}
\end{figure}

In particular, in the thermodynamic limit with $N\to \infty$, we can apply the Holstein-Primakoff transformation
\begin{align}
\hat{J}_z&\xrightarrow{}\hat{b}^{\dagger}\hat{b}-\frac{N}{2}, \\ \hat{J}_{+}&\xrightarrow{}\hat{b}^{\dagger}\sqrt{N-\hat{b}^{\dagger}\hat{b}}, \\ \hat{J}_{-}&\xrightarrow{}\sqrt{N-\hat{b}^{\dagger}\hat{b}}~\hat{b},
\end{align}
such that Hamiltonian (\ref{eq3}) becomes
\begin{align}
\hat{H}_{\text{hp}} &= \Delta_{c}\hat{a}^{\dagger}\hat{a}+(\Delta_{a}+V)\hat{b}^{\dagger}\hat{b}+g(\hat{a}\hat{b}^{\dagger}+\hat{a}^\dagger\hat{b})\nonumber\\
&+\frac{\lambda}{2}(\hat{a}^{\dagger2}+\hat{a}^{2})
-\frac{\Delta_a N}{2}.
\label{eq13}
\end{align}
Defining $\boldsymbol{v}=[\hat{a},\hat{a}^{\dagger},\hat{b},\hat{b}^{\dagger}]^{T}$, one derives its equation of motion as $\dot{\boldsymbol{v}}(t) = M \boldsymbol{v}(t)$, where \begin{align}
  M =
    \begin{pmatrix}
      -(i\Delta_{c}+\kappa)     & -i\lambda &-ig & 0 \\
       i\lambda & i\Delta_c-\kappa  & 0&  ig   \\
       -ig&0&-i(\Delta_{a}+V)&0\\
       0& ig& 0& i(\Delta_{a}+V)
    \end{pmatrix}.
    \label{eq14}
\end{align}
The retarded Green's function is defined as
\begin{equation}
\label{eq:gr_def}
G_{i,j}^{\rm{R}}(t) = -i \left\langle \left[{v_i(t)},{v_j^{\dagger}(0)}\right] \right\rangle \theta(t).
\end{equation}
For convenience, we set $S_{i,j} = \left\langle \left[ v_i(0),\, v_j^{\dagger}(0) \right] \right\rangle$. Using
$\dot{v}_i(t) = M_{i,k} v_k(t)$ and Fourier transforming the expression above, we derive
the inverse of the retarded Green's function~\cite{Adv.QuantumTechnol.2.1800043}
\begin{equation}
    [G^{\mathrm{R}}(\omega)]^{-1} = S^{-1}(\omega - \mathrm{i}M),
    \label{eq15}
\end{equation}
where $S = \mathrm{diag}(1, -1, 1, -1)$ (see Appendix \ref{app:labelA} for details). We then have
\begin{align}
 & [G^{\mathrm{R}}(\omega)]^{-1}=\nonumber\\
&    \begin{pmatrix}
      \omega-\Delta_{c}+i\kappa& -\lambda &-g & 0 \\
       -\lambda &-(\omega+\Delta_c+i\kappa)&0&-g\\
       -g&0&\omega-\Delta_{a}-V&0\\
       0&-g&0&-(\omega+\Delta_{a}+V)
    \end{pmatrix}.
    \label{eq16}
\end{align}
Setting $\rm det|[G^{\mathrm{R}}(\omega=0)]^{-1}|=0$, we have
\begin{equation}
    (\kappa^2-\lambda^2)(\Delta_a+V)^2+\left[g^2-(\Delta_a+V)\Delta_c\right]^2 = 0,
    \label{eq17}
\end{equation}
whose solution leads to the critical coupling for the superradiant transition for $N\to \infty$
\begin{align}
\lambda_c=\sqrt{\kappa^2+\frac{\left[g^2-(\Delta_a+V)\Delta_c\right]^2}{(\Delta_a+V)^2}}.
    \label{eq18}
\end{align}
Here several remarks are in order. First, as implied by Eq.~(\ref{eq17}), the solution to $\rm det|[G^{\mathrm{R}}(\omega=0)]^{-1}|=0$ exists only when $\lambda>\kappa$, meaning superradiance only occurs for $\lambda>\kappa$. This is consistent with the analytical results for $V=0$, as well as our numerical results under finite $N$.
Second, for sufficiently large $V>0$, $\lambda_c$ asymptotically approaches $\sqrt{\kappa^2+\Delta_c^2}$. This is consistent with the observations in Fig.~\ref{fig5}.
Third, similar to previous discussions for $V=0$, the boundaries $\lambda=\kappa$ and $\lambda_c$ in Eq.~(\ref{eq18}) intersect at $\lambda=\kappa$ and $V=-\Delta_a+\frac{g^2}{\Delta_{c}}$.
Hence, for $V<-\Delta_a+\frac{g^2}{\Delta_{c}}$, the phase transition occurs at $\lambda=\kappa$, whereas for $V>-\Delta_a+\frac{g^2}{\Delta_{c}}$, the phase transition occurs at $\lambda_{c}$ in Eq.~({\ref{eq18}}). The latter is consistent with the monotonic trend of the phase boundary toward larger repulsive interactions. Moreover, as previously discussed, for $V<-\Delta_a+\frac{g^2}{\Delta_{c}}$, the boundary of Eq.~({\ref{eq18}}) separates the superradiant phase into two distinct regions. The left contains coexisting superradiant and normal solutions, and the right only support superradiant solutions.

In Fig. \ref{fig7}(b), we plot the phase boundary of Eq.~(\ref{eq18}) in red dashed curve, along with the phase boundary $\lambda=\kappa$ (also in red dashed line). Moreover, $\lambda_{c}$ of Eq.~(\ref{eq18}) diverges to $+\infty$ as $V\to-\Delta_a$ (green dashed line). For $V<-\Delta_a$, $\lambda_{c}>\sqrt{\kappa^2+\Delta_{c}^2}$ which lies out of the range considered in this study. We note that for $V=0$, the resulting expression Eq.~(\ref{eq18}) reduces to the boundary discussed previously. Together, they agree well with the numerically evaluated phase boundary for $N=75$ (color contour in the background).
From numerical simulations, we estimate that the phase boundaries would converge to those in the thermodynamic limit when $N\gtrsim 45$.

Finally, the enhanced superradiant phase and the emergence of several finger-like normal regions above the $\lambda=\kappa$ boundary are found to persist even when realistic dipole–dipole interactions are introduced. The interaction takes the form $V_{\mathrm{r}}=\frac{C_{3}}{R^{3}}$ \cite{ChinesePhys.B30.020305}, leading to the following modified Hamiltonian
\begin{align}
\hat{H}_{\text{r}} &= \Delta_{c}\hat{a}^{\dagger}\hat{a}+\Delta_{a}\sum_{i=1}^N\frac{\sigma_{i}^{z}}{2}+\frac{g}{\sqrt{N}}\sum_{i=1}^N(\hat{a}^{\dagger}\hat{\sigma}_{i}^{-}+\hat{a}\hat{\sigma}_{i}^{+})\nonumber\\
&+\frac{\lambda}{2}(\hat{a}^{\dagger2}+\hat{a}^{2})+\sum_{i<j}\frac{C_{3}}{R^3_{ij}}(\hat{\sigma}_{i}^{+}\hat{\sigma}_{j}^{-}+\text{H.c.}),
\label{eq19}
\end{align}
where $R_{ij} = R_0 |i-j|$, and $R_0$ denotes the interatomic distance. By varying $R_0$, we can tune the strength of the interatomic interaction $V_s = \frac{C_3}{R_0^3}$. 
A phase diagram calculated using typical experimental parameters are shown in Fig.~\ref{fig8}. And the interactions effects are clearly visible, as the phase boundary depends sensitively on the variation of $R_0$.

\section{Conclusion}
We have analyzed in detail the impact of interatomic interactions on the phase boundary of the superradiant phase transition in a parametrically driven, dissipative Tavis–Cummings model. Our results show that the phase boundary is substantially modified by the presence of interatomic interactions.
In the attractive regime, the dipole-dipole interactions qualitatively change the transition behavior: even under weak atom-cavity coupling rates, the superradiance transition can emerge as soon as the gain coefficient surpasses the dissipation rate.
By contrast, this only occurs for very large atom-cavity coupling rates in the absence of dipolar interactions.
This indicates that attractive interactions effectively lower the critical threshold for the emergence of superradiance.
In the repulsive regime, superradiance is strongly suppressed. These results highlight the crucial role of long-range atomic interactions in shaping the phase diagram and provide valuable insights into the controllability of the transition in realistic experimental settings. 

In our configuration, we consider all fields to be co-propagating along the $x$-direction as $\mathrm{TEM}_{00}$ traveling waves, so that the beam waists are sufficiently broad compared to the atomic ensemble, ensuring uniform atom-field coupling strengths.
Such configurations can be implemented using the Rydberg states $\ket{60S_{1/2}, m_{j}=\tfrac{1}{2}}$ and $\ket{60P_{1/2}, m_{j}=-\tfrac{1}{2}}$ of $^{87}\mathrm{Rb}$ as $\ket{s}$ and $\ket{r}$, respectively, to induce the desired atomic interactions. In particular, optical tweezers can be employed to trap and arrange the atomic array, while the nonlinear drive can be generated by an optical parametric amplifier inside the cavity~\cite{Phys.Rev.A80.033807}. All these techniques are currently feasible in laboratories.

\begin{acknowledgments}
This work is supported by the National Natural Science Foundation of China (Grant No. 12374479), and the Innovation Program for Quantum Science and Technology (Grant No. 2021ZD0301205).
\end{acknowledgments}

\appendix
\section{retarded Green's functions}\label{app:labelA}

In this appendix we show how to obtain the retarded Green's function, defined as
\begin{equation}
\label{eq:step1}
G_{i,j}^{\rm{R}}(t) = -i \left\langle \left[{v_i(t)},{v_j^{\dagger}(0)}\right] \right\rangle \theta(t).
\end{equation}
Here $\theta(t)$ is the Heaviside step function ($\theta(t)=1$ for $t\geq0$, $\theta(t)=0$ for $t<0$). For convenience, we set $S_{i,j} = \left\langle \left[ v_i(0),\, v_j^{\dagger}(0) \right] \right\rangle$. By using
$\dot{v}_i(t) = M_{i,k} v_k(t)$, we get the time derivative of the Green's function
\begin{align}
    \partial_t G_{i,j}^{\rm{R}}(t) &= -i \partial_t \left[ \left\langle \left[{v_i(t)},{v_j^\dagger(0)}\right] \right\rangle \theta(t) \right] \nonumber\\
    &= -i \left\langle \left[{\dot{v}_i(t)},{v_j^\dagger(0)}\right] \right\rangle \theta(t) - i \left\langle \left[{v_i(t)}, {v_j^\dagger(0)}\right] \right\rangle \delta(t) \nonumber\\
    &= -i M_{i,k} \left\langle \left[{v_k(t)},{v_j^\dagger(0)}\right] \right\rangle \theta(t) - i \left\langle \left[{v_i(0)},{v_j^\dagger(0)}\right] \right\rangle \delta(t)\nonumber\\
     &= M_{i,k} G_{k,j}^{\rm{R}}(t) - i S_{i,j}\delta(t).
    \label{eq:step2}
\end{align}
We now apply the Fourier transform to Eq.~(\ref{eq:step2}), defined as $\mathcal{F}[f(t)] = \int_{-\infty}^{\infty} dt \, e^{i\omega t} f(t)$, to have
\begin{align}
-i\omega G^{\rm{R}}(\omega) = M G^{\rm{R}}(\omega)-i S.
\end{align}
This finally leads to 
\begin{align}
    [G^{\mathrm{R}}(\omega)]^{-1} = S^{-1}(\omega - \mathrm{i}M).
    \label{eq:gr_inverse_appendix}
\end{align}

\end{document}